\documentclass[10pt]{article}
\usepackage[a4paper,bottom=0.8in,left=0.8in,right=0.8in,top=0.8in]{geometry}
\usepackage{graphicx}
\usepackage{xcolor}
\newcommand{\denetver}{0.10.3}
\newcommand{\benchhost}{AMD EPYC 7742 server}
\usepackage[T1]{fontenc}
\usepackage[utf8]{inputenc}
\usepackage{hyperref}
\hypersetup{
    colorlinks=true,
    linkcolor=blue,
    filecolor=magenta,
    urlcolor=cyan
}
\usepackage{microtype}
\usepackage{inconsolata}
\usepackage{float}
\usepackage{tabularx,booktabs}
\newcommand{\perstepcmidrules}{\cmidrule(lr){2-4} \cmidrule(lr){5-7}}
\newcolumntype{Y}{>{\centering\arraybackslash}X}
\usepackage{orcidlink}
\usepackage[misc]{ifsym}
\usepackage{enumitem}
\usepackage{listings}
\definecolor{codebg}{HTML}{F6F6F4}
\definecolor{codecomment}{HTML}{6A737D}
\definecolor{codekw}{HTML}{005A8B}
\definecolor{codestr}{HTML}{8B2E00}
\definecolor{coderule}{HTML}{B0B0B0}
\title{\texttt{denet}, a lightweight command-line tool for process monitoring in bioinformatics benchmarking and beyond}

\author{Ben Carrillo \orcidlink{0009-0003-5704-4151}, Izaskun Mallona$^{*}$ \orcidlink{0000-0002-2853-7526}\\
\\
Department of Molecular Life Sciences, University of Zurich\\
and SIB Swiss Institute of Bioinformatics, Zurich, Switzerland \\
$^{*}$Present address: Department of Biosystems Science and Engineering,\\
ETH Zurich, Basel, Switzerland \\
{ \Letter: \texttt{ben.uzh@proton.me} and \texttt{izaskun.mallona@bsse.ethz.ch}}
}

\date{}

\begin{document}
\maketitle

\section*{Abstract}

\paragraph{Summary:} \texttt{denet} is a lightweight process monitoring tool providing real-time resource profiling of running processes. It reports CPU, memory, disk I/O, network activity, and thread usage, including recursive child monitoring, with adaptive sampling rates. \texttt{denet} exposes two interfaces: a CLI with colorized output and a Python API. Output format is JSONL. Each record carries performance metrics and process metadata (PID and the executed command). Structured output makes \texttt{denet} suitable for benchmarking and tuning data-intensive pipelines in bioinformatics and beyond. We provide CLI and API examples, including a bioinformatics workflow in Snakemake, showcasing \texttt{denet}'s diagnostic properties.

\paragraph{Availability and implementation:} \texttt{denet} is open-source software released under the GPLv3 terms, and maintained at \url{https://github.com/btraven00/denet}. It is implemented in Rust, with Python bindings provided via \texttt{maturin}, and installs from Cargo (\texttt{cargo install denet}), PyPI (\texttt{pip install denet}) or conda-forge. Most functionality does not require administrative privileges, enabling use on cloud platforms, HPC clusters, and standard Linux workstations.

\vspace{0.5cm}

\noindent\textbf{Keywords:} resource profiling, workflow, benchmarking, bioinformatics

\section*{Introduction}
\label{sec:intro}

Resource usage (CPU, memory, disk and network I/O) shapes bioinformatics tool development, parameter tuning, and workflow design. Standard command-line tools show limitations. System-level profilers (\texttt{top}, \texttt{htop}) lack process specificity and cannot isolate the resource footprint of a single job. Process-summary utilities (\texttt{time}) report cumulative totals only at completion, missing the dynamics that often indicate bottlenecks. Their text output is built for interactive use, which complicates integration into workflow managers.

Workflow managers such as \texttt{make}, \texttt{snakemake}, and \texttt{nextflow} are often coupled to coarse, non-customizable resource profilers. \texttt{denet} grew out of our own benchmarking work, where we repeatedly needed to attribute a resource peak to a particular stage of a workflow and found that neither the aggregate summaries produced by workflow managers nor the interactive system profilers could tell us which stage was responsible. We describe \texttt{denet} (Turkish root for `to inspect' or `to check'), a new process monitoring toolkit with a Unix-style CLI and a Python API. \texttt{denet} adds adaptive sampling, recursive process-tree monitoring, and experimental eBPF-based off-CPU analysis, with easy integration to Snakemake and other workflow tools.

Here we ask two questions: does sub-second profiling of the whole process tree perturb the workflow it measures, and does it support decisions that the aggregate summaries of workflow managers (wall-clock time, peak memory and I/O totals) cannot? We answer the first with measured overhead and the second with a case study.

\section*{Implementation}

\subsection*{Adaptive profiling with user-provided resolution}

\texttt{denet} samples adaptively between a user-set base interval and a maximum interval, in milliseconds (100 and 1000~ms by default); setting both equal gives a fixed rate. During the first second, \texttt{denet} samples at the base interval. Over the next nine seconds, the interval grows linearly to the maximum, and stays there for processes running over 10~s. Adaptive sampling resolves short-lived or dynamic tasks at high resolution while keeping system overhead low for long-running ones.

\subsection*{Metrics}

\texttt{denet} collects: CPU usage, per process and aggregated over the tree (\texttt{top}/\texttt{htop} convention, 100\% = one fully loaded core); memory, separating RSS and VMS to expose peak usage and possible memory leaks, and major page faults (including swap-ins); GPU memory and percent usage (requires the NVIDIA Management Library NVML); disk reads and writes, in cumulative bytes; network bytes received and sent (TCP per process tree, otherwise system-wide); thread counts and child processes spawned from a parent, giving insight into concurrency. As summary metadata, \texttt{denet} reports the full command of the monitored process, the path to its executable, PID, runtime duration, and the metric sources available on the host. In \texttt{run} mode, \texttt{denet} exits with the monitored command's exit status. Tracking of child processes can be disabled with \texttt{--exclude-children}.

\subsection*{eBPF support and off-CPU analysis}

On GNU/Linux, \texttt{denet} offers experimental support for the extended Berkeley Packet Filter (eBPF), a native Linux kernel technology that traces events with low overhead by aggregating them in the kernel and passing only summaries to user space \cite{gregg2019bpf}. \texttt{denet} attaches eBPF programs to system calls, network functions and the scheduler, giving per-process visibility into system calls, network traffic and off-CPU time.

eBPF observes processes from the kernel, including those inside containers (e.g., Docker, Apptainer), common in bioinformatics workflows, so \texttt{denet} can profile them from the host without modifying the application. From the eBPF traces \texttt{denet} can also quantify off-CPU time; that reveals where the profiled application is waiting and points to bottlenecks such as I/O and locks.

\subsection*{Architecture}

\texttt{denet} samples the monitored process and its descendants, re-walking the tree at every sample so that processes spawned mid-run are included, and opportunistically integrates multiple sources of information potentially exposed by the host: \texttt{/proc} and the \texttt{sysinfo} crate for CPU, memory and I/O; hardware counters (IPC, last-level cache misses, backend stalls), pressure stall information and RAPL energy where the kernel permits; NVML for GPU metrics; and eBPF probes through the Aya crate for off-CPU time, syscalls and network bytes. Samples are aggregated into process-tree records and written as JSONL while the run proceeds, to be consumed by the CLI, the Python API, or the report generator. A data-flow diagram and a description of each stage are maintained with the code. The CLI and Python API are described in Supplementary Notes~2 and~3, respectively.

\subsection*{Comparison to other tools}

Unlike \texttt{top}/\texttt{htop}, \texttt{ps}, \texttt{time}, \texttt{pidstat}
and \texttt{psutil}/\texttt{psrecord} \cite{psrecord,rodola2020psutil}
(Supplementary Table~S1), \texttt{denet} samples adaptively. It also combines an
API, recursive process-tree tracking, structured streaming output, eBPF and GPU
metrics. \texttt{bpftrace} \cite{bpftrace} shares the tree tracking and eBPF but
is an interactive tracing language; \texttt{bpftime} \cite{zheng2025} is an eBPF
runtime, not a profiler. Three tools are adjacent rather than overlapping.
LIKWID \cite{treibig2010likwid} reads hardware counters on CPUs and NVIDIA GPUs,
but per CPU rather than per process, and per-region attribution requires
instrumenting the application. \texttt{hyperfine} \cite{hyperfine} reports the
distribution of wall-clock time over repeated runs, with no resource metrics.
\texttt{perf} samples user and kernel call stacks, rendered as flame graphs
\cite{gregg2016flame}: a richer view of where CPU time goes, but it needs
privileges for kernel stacks, symbols and expertise, and does not follow memory
or I/O over time.

\section*{Application: profiling DNA mapping and variant calling}

Suppose we need to make a workflow cheaper to run. We use aggregate summaries on a DNA mapping and variant calling workflow to identify candidates for optimization. After ensuring our profiling does not perturb the pipeline, we made two rounds of changes and re-profiled: together they halved peak memory and cut wall time by two-thirds on the target rule, with identical alignments and duplicates. We use this example to illustrate a methodology and not as a performance claim about bioinformatics workflows in general.

\subsection*{Workflow}

We simulate a genome and paired-end reads with \texttt{wgsim} \cite{wgsim}, index the genome and align the reads with \texttt{bowtie2} \cite{langmead2012fast}, sort, index and mark duplicates with \texttt{samtools} \cite{li2009sequence}, and call variants with \texttt{bcftools} (full list in Supplementary Note~5). The simulated genome is 50~Mb (five chromosomes of 10~Mb), covered 30-fold by five million paired-end 150~nt reads. A further 20\% of the pairs were then copied and added, giving six million pairs at 36-fold coverage. At this depth the duplicates that marking finds are the injected copies, as in a real library, rather than chance collisions from oversampling a small genome.  Each rule was run three times.

\subsection*{Profiling and overhead}

We integrated \texttt{denet} with the Snakemake \cite{koster2012snakemake} workflow in two ways. First, in \emph{wrap} mode: each rule's command is passed to \texttt{denet run} (one helper in the Snakefile wraps every rule's command), which records a JSONL with the time series and propagates exit status to Snakemake. Second, and as a reference only, in \emph{native} mode, \texttt{denet}'s Python API patches the internal Snakemake sampler at runtime, so the reported aggregates are measured by from \texttt{denet} without changes to the rules. This mode observes peaks that Snakemake's sampler misses (Supplementary Table~S3).

We collected baseline data from \texttt{psutil-based} stats in Snakemake, (\emph{baseline}), \texttt{denet} wrap, and \texttt{denet} native, on identical hardware and pre-built software environments: \texttt{denet} \denetver{} on an \benchhost{} (Ubuntu 24.04).

Measurement overhead was small (Table~S2). Summed over rules, wall-clock time was 1780~s at baseline, 1785~s in wrap mode and 1786~s in native mode ($+0.3$--$0.4$\%); both modes add 0.1--0.45~s to short rules and stay within run-to-run noise on long ones, and the \texttt{denet} process itself about 6~MB of memory (Supplementary Note~7). The sampler's own cost matters more when many jobs run at once: with 200 concurrent 1.6~s jobs on 32 cores, Snakemake's \texttt{psutil} sampler slowed the run to $30.5 \pm 0.3$~s and recorded no peak memory for any job, against $18.5 \pm 0.2$~s with \texttt{denet} native, which recorded all 200. Monitoring does not change the results: with seeded simulation, checksums of the reads, alignments, duplicates and variant calls are identical in all three conditions.

\subsection*{Optimization target}

In our run, duplicate marking needed 3.2~GB, four times any other rule (Supplementary Table~S2 and Supplementary Figure~S1B). This rule sets the memory the whole workflow must request; yet it kept only about two cores busy, although each of its four stages had two threads. Alignment is the longest stage, but it uses four cores with steady memory.

\subsection*{Diagnosis}

The target rule pipes \texttt{samtools sort -n}, \texttt{fixmate}, a coordinate \texttt{sort} and \texttt{markdup} (a form in current use~\cite{rat2026pangenome,pnas2026apobec}). \texttt{denet}'s per-child records locate the 3.2~GB memory peak (Figure~\ref{fig:benchmark}, left). The name sort reads its entire input, filling its memory allowance and spilling the rest to disk; for the first 22~s the downstream processes are idle. While it merges and emits, it keeps that allowance, and the coordinate sort fills and spills a second: the peak can be attributed to two distinct sort buffers, each at \texttt{samtools}' default limit, held at once. Setting each sort's memory limit independently confirms this: each sort's own peak follows its own limit only, and the rule's peak is their sum. The drop at 78~s is the name sort exiting. Both spills appear in \texttt{denet}'s per-process disk writes (374 and 542~MB) and in \texttt{samtools}' own log messages, aligned with the trace. Each of the coordinate sort's three spills also stalls the pipe upstream: while it sorts and writes a buffer it stops reading, and \texttt{fixmate} and the name sort drop from over 100\% to under 10\% CPU until it resumes (Figure~\ref{fig:benchmark}, left, B--C).

\subsection*{Remediation}

\begin{figure}[H]
  \centering
  \includegraphics[width=\linewidth]{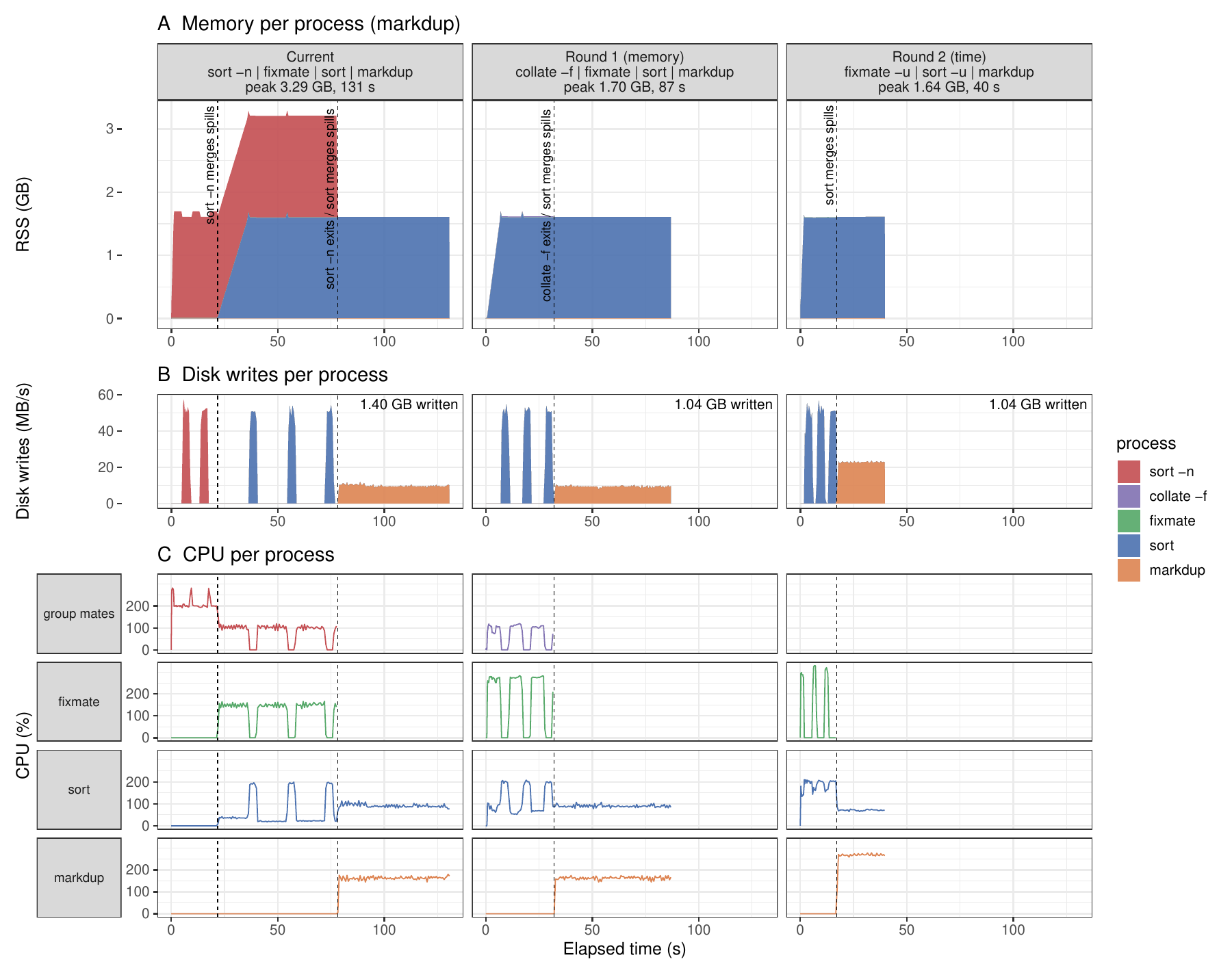}
  \caption{Per-process profiling of the duplicate-marking rule, 30-fold simulated coverage, one repeat of each version, recorded by \texttt{denet} in wrap mode. Columns: the current pipeline; round 1, \texttt{samtools collate -f} in place of the name sort, which removes one of the two sort buffers and shortens the rule; round 2, uncompressed pipes with no grouping step, which removes redundant compression. \textbf{A}, resident memory stacked by process, with each version's peak and duration in its header; \texttt{collate -f}, \texttt{fixmate} and \texttt{markdup} stay under 15~MB and do not show at this scale. \textbf{B}, disk write rate stacked by process, with the total written: each burst from a sort is a full memory buffer spilled to a temporary file, and \texttt{markdup} writes the output. \textbf{C}, CPU per process (100\% = one core), one lane per pipeline position. Dashed lines are events logged by the pipeline itself on the same clock as \texttt{denet}'s samples. All axes linear. }
  \label{fig:benchmark}
\end{figure}

Limiting memory in either sort would address only half of the peak, whereas removing the name sort removes an entire buffer. \texttt{fixmate} requires mates only to be adjacent, not name-ordered, which \texttt{samtools collate -f} provides with bounded memory, which is our case because \texttt{bowtie2} reports one alignment per read. This is the pipeline the \texttt{samtools} documentation prescribes for duplicate marking, noting that no additional sort steps are normally needed~\cite{samtoolsmarkdup}. This substitution reduced peak memory from $3306 \pm 0.4$ to $1719 \pm 43$~MB ($-48$\%), wall time from $131.6 \pm 0.5$ to $87.1 \pm 0.5$~s ($-34$\%) and the disk writes \texttt{denet} records from 1.40 to 1.04~GB, flagging the same duplicate reads, with identical alignments (Figure~\ref{fig:benchmark}, middle).

The revised trace pointed to a second change: \texttt{fixmate} ran at 170--280\% CPU for light work, mostly compressing data that the next stage decompressed at once. Uncompressed pipes (\texttt{-u}), without the grouping step that \texttt{bowtie2}'s adjacent mates make unnecessary, cut wall time to $40.2 \pm 0.1$~s and CPU time from $201 \pm 1$ to $51 \pm 0.5$~s, again with identical alignments and duplicates, and unchanged memory, now limited by the coordinate sort (Figure~\ref{fig:benchmark}, right). Omitting the grouping step is documented as safe for a file already name-ordered or collated~\cite{samtoolsmarkdup}. A comparison of the two changes, with instruction counts and off-CPU time from eBPF, helps attributing most of the gain to the uncompressed pipes. Aggregate summaries can tell which pipeline is cheaper, not why: each change here rested on attributing memory and CPU to a specific process at a specific time.

\section*{Conclusion}

\texttt{denet} is a lightweight and customizable process monitoring tool built for workflows and benchmarks. It adds adaptive sampling, recursive process-tree monitoring, and eBPF-based off-CPU analysis. Its dual interface (CLI for interactive use and POSIX workflows, Python API for scripting) fits both settings. On a Snakemake workflow, \texttt{denet} does not perturb the default profiling readouts, and where its readings differ from them it is on brief peaks, which its finer sampling catches more often. Its joint CPU, memory, and I/O timeseries attribute resource peaks to specific pipeline stages and to individual child processes within them. The resulting changes reduced peak memory by 48\% and then wall time by a further half, without changing the alignments or the duplicates found. \texttt{denet} is Linux-first. macOS support via host APIs is experimental, and eBPF needs Linux 5.8 or newer (or root on older kernels) with \texttt{CONFIG\_BPF\_SYSCALL}.

\section*{Competing interests}

The authors declare no competing interests.

\section*{Data and code availability}

\texttt{denet} is available at \url{https://github.com/btraven00/denet} and the example workflow at  \url{https://github.com/btraven00/denet_example}, both under a GPLv3 license. The results reported here were produced with \texttt{denet} v\denetver{} and \texttt{denet\_example} v1.0.0; the measurement outputs of that run are archived on Zenodo (\url{https://doi.org/10.5281/zenodo.23105224}). The workflow pins that \texttt{denet} version in its \texttt{conda} environment. Supplementary material (CLI and Python API usage, extended comparison table, installation notes) describes the package and the workflow in detail.

\section*{Funding}

This project received no specific funding. IM acknowledges the University of Zurich GRC Career Grant 2025\_Q1\_CG\_001.

\section*{Acknowledgements}

We thank Mark D. Robinson, Charlotte Soneson, Daniel Incicau and Atreya Choudhury for comments on the manuscript.

\section*{Authors' contributions}

BC and IM conceived the study. BC designed and implemented \texttt{denet}. IM created the profiling workflow; BC extended it for this revision. IM wrote the original draft; BC wrote the revised manuscript.

\bibliographystyle{unsrt}
\bibliography{refs}

\begin{thebibliography}{10}

\bibitem{gregg2019bpf}
Brendan Gregg.
\newblock {\em BPF performance tools}.
\newblock Addison-Wesley Professional, 2019.

\bibitem{psrecord}
Thomas Robitaille.
\newblock \texttt{psrecord}: Record the cpu and memory activity of a process.
\newblock \url{https://github.com/astrofrog/psrecord}, 2025.
\newblock Version 1.4.

\bibitem{rodola2020psutil}
Giampaolo Rodola.
\newblock \texttt{psutil} documentation.
\newblock {\em Psutil. \url{https://psutil.readthedocs.io/en/latest}}, 2025.

\bibitem{bpftrace}
bpftrace contributors.
\newblock bpftrace: High-level tracing language for linux ebpf.
\newblock \url{https://github.com/bpftrace/bpftrace}, 2025.
\newblock Accessed: 2025-09-11.

\bibitem{zheng2025}
Y.~Zheng, T.~Yu, Y.~Yang, Y.~Hu, X.~Lai, D.~Williams, and A.~Quinn.
\newblock Extending applications safely and efficiently.
\newblock In {\em Proceedings of the 19th USENIX Symposium on Operating Systems
  Design and Implementation (OSDI 25)}, pages 557--574, 2025.

\bibitem{treibig2010likwid}
Jan Treibig, Georg Hager, and Gerhard Wellein.
\newblock {LIKWID}: A lightweight performance-oriented tool suite for x86
  multicore environments.
\newblock In {\em 39th International Conference on Parallel Processing
  Workshops}, pages 207--216, 2010.

\bibitem{hyperfine}
David Peter.
\newblock \texttt{hyperfine}: A command-line benchmarking tool.
\newblock \url{https://github.com/sharkdp/hyperfine}, 2018.

\bibitem{gregg2016flame}
Brendan Gregg.
\newblock The flame graph.
\newblock {\em Communications of the ACM}, 59(6):48--57, 2016.

\bibitem{wgsim}
Heng Li.
\newblock wgsim: Read simulator.
\newblock \url{https://github.com/lh3/wgsim}, 2011.
\newblock Accessed: 2026-04-08.

\bibitem{langmead2012fast}
Ben Langmead and Steven~L Salzberg.
\newblock Fast gapped-read alignment with bowtie 2.
\newblock {\em Nature methods}, 9(4):357--359, 2012.

\bibitem{li2009sequence}
Heng Li, Bob Handsaker, Alec Wysoker, Tim Fennell, Jue Ruan, Nils Homer, Gabor
  Marth, Goncalo Abecasis, Richard Durbin, and 1000 Genome Project
  Data~Processing Subgroup.
\newblock The sequence alignment/map format and samtools.
\newblock {\em bioinformatics}, 25(16):2078--2079, 2009.

\bibitem{koster2012snakemake}
Johannes K{\"o}ster and Sven Rahmann.
\newblock Snakemake—a scalable bioinformatics workflow engine.
\newblock {\em Bioinformatics}, 28(19):2520--2522, 2012.

\bibitem{rat2026pangenome}
Kai Li, Julia~L. Ciosek, Sergey Koren, Adam~M. Phillippy, Yaming Zhu,
  William~A. Lauer, et~al.
\newblock Telomere-to-telomere genome assembly and a pangenome for the rat.
\newblock {\em Cell Genomics}, 6:101281, 2026.

\bibitem{pnas2026apobec}
Alexandra Puzon, Atri Raval, Cameron Cordero, Daniel Schiefen, Piotr~A.
  Mieczkowski, Steven~A. Roberts, et~al.
\newblock Distinct repair processes produce apobec-induced deletions, tandem
  substitutions, and complex mutations in yeast and human cells.
\newblock {\em Proceedings of the National Academy of Sciences},
  123:e2609794123, 2026.

\bibitem{samtoolsmarkdup}
The SAMtools/HTSlib project.
\newblock {\em samtools markdup manual page}, 2026.
\newblock \url{http://www.htslib.org/doc/samtools-markdup.html}.

\end{thebibliography}

\clearpage
\setcounter{section}{0}
\renewcommand{\thesection}{\arabic{section}}
\setcounter{table}{0}
\renewcommand{\thetable}{S\arabic{table}}
\setcounter{figure}{0}
\renewcommand{\thefigure}{S\arabic{figure}}

\begin{center}{\large\bfseries Supplementary material}\end{center}
\vspace{0.5em}

\section{Installation}
\label{sec:install}

\texttt{denet} can be installed as a Rust crate or as a Python package. Additionally, conda packages and a Nix flake are available:

\begin{lstlisting}[language=bash]
# From the Rust ecosystem (recommended for the CLI):
cargo install denet

# From PyPI (Python API and denet-report; no CLI):
pip install denet

# From conda-forge:
conda install -c conda-forge denet

# with pixi:
pixi add denet

# with Nix (flakes), from the repository; builds the CLI without eBPF.
# Append /v0.10.3 to the reference to pin a release:
nix profile install github:btraven00/denet
\end{lstlisting}

Most functionality runs as an unprivileged user. The eBPF backend requires a Linux kernel compiled with \texttt{CONFIG\_BPF\_SYSCALL} and \texttt{CAP\_BPF}, \texttt{CAP\_PERFMON} and \texttt{CAP\_DAC\_READ\_SEARCH} (Linux 5.8 or newer) or root, and a build that includes it: the Linux wheels on PyPI do, through the Python API; \texttt{cargo install} needs \texttt{--features ebpf} (and \texttt{clang} at build time); the conda-forge package and the Nix flake currently ship without it. The NVIDIA GPU backend requires the NVIDIA Management Library (NVML), but no elevated privileges.

\section{Command-line interface}
\label{sec:cli}

\begin{lstlisting}[language=bash]
# Real-time monitoring of the process `sleep 5` (to the terminal):
denet run sleep 5

# Generate a report in JSON format, including metadata on the first line:
denet --json run sleep 5 > metrics.json

# Set the sampling interval (in milliseconds):
denet --interval 500 run sleep 5

# Set the upper bound of the adaptive sampling interval:
denet --max-interval 2000 run sleep 5

# Monitor an existing running process with PID 1234:
denet attach 1234

# Monitor PID 1234 for 10 seconds only:
denet --duration 10 attach 1234

# Write to a file only, with no terminal output:
denet --quiet --json --out metrics.jsonl run python script.py

# Disable child process monitoring (only track the parent process):
denet --exclude-children run python multi_process_script.py

# Summarise a previously saved metrics file:
denet stats metrics.jsonl

# Render CPU, memory and I/O plots from a saved run
# needs to install the python extra with pip:
# pip install denet[report]:
denet-report metrics.jsonl -o report.html
\end{lstlisting}

\section{Python API}
\label{sec:api}

\begin{lstlisting}[language=python]
#!/usr/bin/python3
import json
import denet

# Create a monitor for a process (command)
monitor = denet.ProcessMonitor(
    cmd=["python", "-c", "import time; time.sleep(10)"],
    base_interval_ms=100,    # Start sampling every 100 ms
    max_interval_ms=1000,    # Sample at most every 1000 ms
    store_in_memory=True,    # Keep samples in memory
    output_file=None,        # Optional file output
    quiet=True,              # Do not print samples to stdout
    include_children=True    # Monitor child processes (default True)
)

# Block until the process exits (other Python threads keep running)
monitor.run()

# All samples, after the process has exited
samples = monitor.get_samples()
print(f"Collected {len(samples)} samples")

# Get summary statistics
summary_json = monitor.get_summary()
\end{lstlisting}

\section{Comparison to related tools}

\begin{table}[H]
\caption{Capabilities across process monitoring tools relevant to bioinformatics benchmarking.  yes: supported; partial: limited or not by default; --: absent. The first column gives what each tool measures; for tools that measure something else, a dash marks a feature outside their scope.}

\centering
\footnotesize
\begin{tabularx}{\linewidth}{l>{\raggedright\arraybackslash}p{3.1cm}YYYYYY}
\toprule
 & Primary measurement & Adaptive sampling & Program\-matic API & Recursive child tracking & Structured output & eBPF backend & GPU observability \\
\midrule
\texttt{top} / \texttt{htop}   & live system-wide process list & --       & --     & partial & --       & --      & partial (nvtop) \\
\texttt{ps}                    & process state at one instant & --       & --     & --      & yes      & --      & --      \\
\texttt{time}                  & totals for one command at exit & --       & --     & --      & yes      & --      & --      \\
\texttt{pidstat}               & periodic per-process counters & --       & --     & partial & yes      & --      & --      \\
\texttt{psutil} / \texttt{psrecord} & per-process sampling from a library & --  & yes    & partial & yes      & --      & --      \\
\texttt{bpftrace}              & ad hoc kernel event tracing & --       & --     & yes     & --       & yes     & --      \\
\texttt{bpftime}               & userspace eBPF runtime & --       & yes    & yes     & yes      & yes     & yes     \\
LIKWID                   & hardware counters per CPU & -- & yes & -- (per CPU) & yes & -- & yes \\
\texttt{hyperfine}       & wall-time distribution over repeated runs & -- & -- & -- & yes & -- & -- \\
\texttt{perf} + flame graphs & sampled call stacks: CPU time per function & -- & partial & yes & partial & -- & -- \\
\texttt{denet}                 & resource time series of a process tree & yes      & yes    & yes     & yes      & yes     & yes     \\
\bottomrule
\end{tabularx}

\label{tab:compare}
\end{table}

\section{DNA alignment and variant calling workflow details}
\label{sec:workflow}

The Snakemake workflow used in the main text is available at \url{https://github.com/btraven00/denet_example}. It comprises ten rules:

\begin{enumerate}[itemsep=0.1pt,topsep=0.5pt]
  \item \texttt{simulate\_genome}: generate a five-chromosome, 50~Mb reference;
  \item \texttt{index\_genome}: \texttt{bowtie2-build};
  \item \texttt{simulate\_reads}: \texttt{wgsim}, five million paired-end 150~nt reads; a random 20\% of the pairs is then copied, under new names that keep each copy paired, giving six million pairs at 36-fold coverage and a 16.7\% duplicate rate;
  \item \texttt{align}: \texttt{bowtie2} paired-end, four threads, piped into \texttt{samtools view};
  \item \texttt{sort\_bam}: \texttt{samtools sort};
  \item \texttt{index\_bam}: \texttt{samtools index};
  \item \texttt{markdup}: \texttt{samtools sort -n} $\vert$ \texttt{samtools fixmate -m} $\vert$ \texttt{samtools sort} $\vert$ \texttt{samtools markdup}, two threads; the \texttt{collate} and uncompressed-pipe variants of the main text are selected with the \texttt{markdup\_group} and \texttt{markdup\_uncompressed\_pipes} configuration keys;
  \item \texttt{index\_markdup}: \texttt{samtools index};
  \item \texttt{faidx\_genome}: \texttt{samtools faidx};
  \item \texttt{call\_variants}: \texttt{bcftools mpileup} $\vert$ \texttt{bcftools call -mv -Oz}, two threads.
\end{enumerate}

Each rule carried a Snakemake \texttt{benchmark} directive with three repeats. Conda-managed environments fixed the software versions.  Environments were built beforehand, outside the benchmark, and activated once rather than per job: per-job activation starts a 90~MB process that would dominate the peak RSS of short rules.

\subsection{Running the workflow}

The execution is wrapped inside a \texttt{Makefile} (\texttt{make paper}); tool environments are built once and put on \texttt{PATH}. This includes:

\begin{lstlisting}[language=bash]
# baseline: plain Snakemake (using a built-in profiler)
snakemake --cores 16 --forceall \
  --config use_denet=false outdir=results_baseline

# denet wrap: each rule's command runs under `denet run`
snakemake --cores 16 --forceall \
  --config use_denet=true outdir=results_denet

# denet native: denet replaces Snakemake's benchmark sampler
snakemake --cores 16 --forceall \
  --config use_denet_native=true outdir=results_denet_native
\end{lstlisting}

The \texttt{denet} wrap condition samples each rule at 50~ms base interval with a 500~ms maximum (\texttt{-i 50 -m 500}), writing a JSONL timeseries to \texttt{results\_denet/denet\_metrics/<step>.jsonl}.

\section{Benchmark summaries}

\begin{table}[H]
\caption{Per-step (coarse) summary statistics across the three benchmarking conditions (three benchmark repeats per rule). Numbers are taken from the Snakemake \texttt{benchmark} directive.  Workflow at 30-fold simulated coverage (50~Mb genome, five million read pairs, 20\% injected duplicates), \texttt{denet} \denetver, Snakemake 9.27.0, \benchhost{}, pinned to one NUMA node; the host was 98\% idle on average during the run. Memory and I/O in binary units ($1$~MB $= 2^{20}$~bytes), as Snakemake and \texttt{denet} report them. In wrap mode, Snakemake's sampler cannot read the memory of the \texttt{denet} process, which holds file capabilities, so the peak RSS shown is the command's alone; \texttt{denet} itself adds about 6~MB (Supplementary Note~\ref{sec:footprint}). On the two sorting rules the reported peak varies between repeats (\texttt{sort\_bam}, 837--872~MB; \texttt{markdup}, 3288--3379~MB), because a brief peak is caught in some repeats only (Supplementary Note~\ref{sec:sampling}). \textsuperscript{a}Snakemake recorded no RSS for one of the three repeats; \textsuperscript{b}for none: the rule ends before Snakemake's first sample.}
\label{tab:perstep}

\centering
\small
\begin{tabular}{lrrrrrr}
\toprule
 & \multicolumn{3}{c}{Wall time (s, mean $\pm$ SD)} & \multicolumn{3}{c}{Peak RSS (MB, mean $\pm$ SD)} \\
\perstepcmidrules
Step & baseline & wrap & native & baseline & wrap & native \\
\midrule
simulate\_genome & 3.49 $\pm$ 0.06 & 3.64 $\pm$ 0.09 & 3.81 $\pm$ 0.13 & 104.1 $\pm$ 12.2 & 97.1 $\pm$ 12.0 & 108.8 $\pm$ 3.1 \\
index\_genome & 41.12 $\pm$ 0.02 & 41.21 $\pm$ 0.11 & 41.56 $\pm$ 0.28 & 143.2 $\pm$ 5.9 & 141.7 $\pm$ 3.4 & 161.6 $\pm$ 0.0 \\
simulate\_reads & 1007.27 $\pm$ 0.54 & 1008.42 $\pm$ 1.89 & 1008.72 $\pm$ 0.99 & 186.6 $\pm$ 0.0 & 186.6 $\pm$ 0.1 & 186.5 $\pm$ 0.1 \\
align & 376.30 $\pm$ 0.28 & 377.97 $\pm$ 1.23 & 378.75 $\pm$ 0.71 & 139.9 $\pm$ 4.6 & 139.8 $\pm$ 4.5 & 141.0 $\pm$ 5.7 \\
sort\_bam & 61.52 $\pm$ 0.13 & 61.82 $\pm$ 0.25 & 61.91 $\pm$ 0.25 & 836.7 $\pm$ 0.0 & 858.9 $\pm$ 19.2 & 870.8 $\pm$ 1.3 \\
index\_bam & 3.27 $\pm$ 0.01 & 3.40 $\pm$ 0.00 & 3.58 $\pm$ 0.06 & 7.2 $\pm$ 0.1 & 7.1 $\pm$ 0.0 & 7.2 $\pm$ 0.1 \\
markdup & 133.32 $\pm$ 1.70 & 132.28 $\pm$ 2.52 & 132.70 $\pm$ 0.34 & 3289.0 $\pm$ 1.6 & 3289.0 $\pm$ 1.7 & 3320.0 $\pm$ 50.7 \\
index\_markdup & 3.42 $\pm$ 0.01 & 3.55 $\pm$ 0.01 & 3.87 $\pm$ 0.01 & 7.2 $\pm$ 0.1 & 7.2 $\pm$ 0.1 & 7.3 $\pm$ 0.0 \\
faidx\_genome & 0.18 $\pm$ 0.04 & 0.30 $\pm$ 0.02 & 0.42 $\pm$ 0.02 & 5.2 $\pm$ 2.2\textsuperscript{a} & --\textsuperscript{b} & 6.7 $\pm$ 0.1 \\
call\_variants & 150.23 $\pm$ 0.33 & 152.12 $\pm$ 0.66 & 151.10 $\pm$ 0.59 & 74.0 $\pm$ 0.0 & 73.9 $\pm$ 0.1 & 74.0 $\pm$ 0.1 \\
\bottomrule
\end{tabular}

\end{table}

The larger wrap-mode overhead on the short rules (\texttt{simulate\_genome}, \texttt{index\_bam}, \texttt{index\_markdup}, \texttt{faidx\_genome}) reflects the fixed cost of spawning a separate \texttt{denet run} process per rule, which is amortised away once rule runtime exceeds a few seconds.

\begin{figure}[H]
  \centering
  \includegraphics[width=\linewidth]{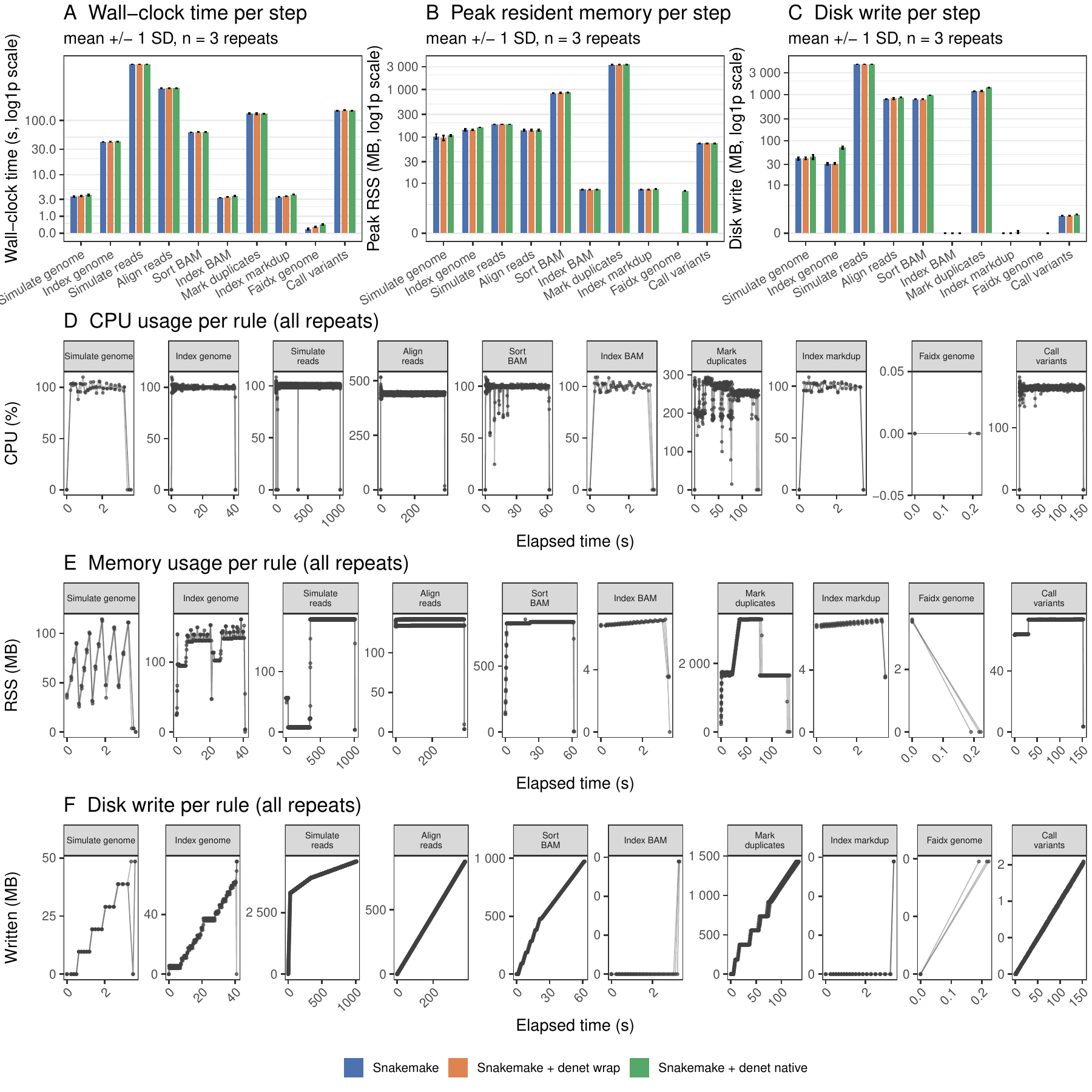}
  \caption{Resource profiling of the whole workflow at 30-fold simulated coverage, three conditions, three repeats per rule. \textbf{A--C}, mean wall-clock time, peak RSS and disk bytes written per rule ($\pm$1~SD; $\log_{10}(1+x)$ axes). \textbf{D--F}, CPU, RSS and cumulative disk-write time series recorded by \texttt{denet} in wrap mode, all repeats, one panel per rule in pipeline order; linear axes.}
  \label{fig:overview}
\end{figure}

\textbf{Supplementary Figure~\ref{fig:overview}} shows the per-rule summaries and the wrap-mode time series for the whole workflow at the same scale.

\section{Memory footprint of \texttt{denet}}
\label{sec:footprint}

In wrap mode the \texttt{denet} process runs alongside the command and adds its own memory. Its peak resident memory (\texttt{denet} \denetver, three repeats) is 5.7~MB for a plain \texttt{denet run} on the \benchhost{} (6.6~MB on a laptop), 22.2~MB with eBPF, for the loaded programs and maps, and 25.6~MB with GPU monitoring (\texttt{--gpu}, laptop with an NVIDIA GPU), for the NVML library; both are opt-in for that reason. Snakemake reports a job holding 50~MiB at 63.1~MB bare and 68.9~MB wrapped ($+5.7$~MB). With file capabilities, which eBPF, hardware counters and RAPL energy need, \texttt{denet} is non-dumpable: its memory map is readable only by root, Snakemake's sampler skips it, and the wrapped job is reported at 63.2~MB. This applies to the run in Table~\ref{tab:perstep}, whose wrap-mode peaks are the command's alone. Scripts: \texttt{validation/footprint/} in the workflow repository.

\section{Sampling interval and peak capture}
\label{sec:sampling}

A sampler reports the largest value it observes, not the largest that occurred. To measure what each sampler misses, each job of a test workflow waits a random 0.5--3~s, so that it starts at a different time relative to the samples, holds 256~MiB of resident memory for a duration $D$, and exits. Each job is measured at once by Snakemake's \texttt{psutil} sampler, by \texttt{denet} native (50~ms, relaxing to 500~ms after 10~s) and by a 10~ms \texttt{psutil} reference. Capture is the reported peak above the interpreter's 13~MiB, as a percentage of the spike (Table~\ref{tab:ladder}).

\begin{table}[H]
\caption{Peak capture by spike duration: mean over eight repetitions, and in brackets the repetitions recovering more than half of the spike. \texttt{denet} \denetver{}, Snakemake 9.27.0, one job at a time, \benchhost{} pinned to one NUMA node. Upper block: random 0.5--3~s delay before the spike; lower block: 20--50~s, once Snakemake samples every 30~s.}
\label{tab:ladder}
\centering
\small
\begin{tabular}{rrrr}
\toprule
Spike duration & 10~ms reference & \texttt{denet} native & Snakemake (\texttt{psutil}) \\
\midrule
0.05~s & 98\% (8/8) & 68\% (5/8) & 34\% (3/8) \\
0.10~s & 100\% (8/8) & 87\% (7/8) & 34\% (2/8) \\
0.25~s & 100\% (8/8) & 100\% (8/8) & 49\% (4/8) \\
0.50~s & 100\% (8/8) & 100\% (8/8) & 92\% (8/8) \\
0.75--2.0~s & 100\% (24/24) & 100\% (24/24) & 100\% (24/24) \\
\midrule
5~s & 100\% (8/8) & 100\% (8/8) & 25\% (2/8) \\
15~s & 100\% (8/8) & 100\% (8/8) & 63\% (5/8) \\
60~s & 100\% (8/8) & 100\% (8/8) & 100\% (8/8) \\
\bottomrule
\end{tabular}
\end{table}

Below Snakemake's 0.5~s interval, spikes were missed in half or more of the repetitions, and whether one was caught depended only on when it started relative to the samples: most repetitions recorded nearly all of the spike or none of it. A sampler with interval $I$ is certain to see a spike of duration $D$ only if $I \le D$. This has four consequences.

First, repeating a benchmark does not recover a missed peak: each repeat catches all of it or none, so the average understates it, and repeating only raises the chance that one run sees it. A fixed start does not help: with the delay fixed at 1.0~s, Snakemake caught a 0.25~s spike in all eight repetitions (4 of 8 with random delays) but 0.05 and 0.1~s spikes in only 3 and 6.

Second, Snakemake's interval grows to 30~s after about 15~s. With the spike delayed by 20--50~s (lower block), it caught a 5~s spike in 2 of 8 repetitions and a 15~s spike in 5, while \texttt{denet}, at 500~ms, caught all. The duplicate-marking peak in the main text lasts about 44~s, longer than 30~s, so that comparison is unaffected.

Third, recovering a spike's shape, not just its peak, needs several samples within it.

Fourth, the same limit applies to cumulative counters. Both samplers report \texttt{/proc} I/O totals as of their last sample, so bytes written after it are lost; reading \texttt{denet}'s trace at Snakemake's sample times reproduces Snakemake's disk totals to within 3\%. The loss grows as a rule gets faster: round 2 of the case study finishes in 40~s, so Snakemake's last sample is at about 15~s and it reports 564~MB of writes against \texttt{denet}'s 1063~MB. Speeding up a rule can thus lower the I/O total that would be used to check it. \texttt{denet} records the time between its last sample and process exit, which bounds what was missed.

Scripts: \texttt{validation/calib/} and \texttt{validation/truncation/} in the workflow repository; \texttt{scripts/check\_truncation.py} reproduces the disk-total comparison, and \texttt{validation/README.md} describes all accuracy checks.

\end{document}